\documentclass[a4paper]{article}
\usepackage[ansinew]{inputenc}
\usepackage{times}
\usepackage[T1]{fontenc}
\usepackage{graphicx}
\usepackage{geometry}
\usepackage{amssymb}
\usepackage{amsmath}
\newcommand{\re}[1]{\textcolor{red}{#1}}
\newcommand{\bl}[1]{\textcolor{blue}{#1}}
\newcommand{\gr}[1]{\textcolor{green}{#1}}
\usepackage{rotating}

\usepackage{tcolorbox}
\usepackage{xcolor}
\colorlet{shadecolor}{gray!25}
\usepackage{subfig}
\usepackage{float}

\author{\normalsize Ludwig A. Hothorn,\\ 
\footnotesize Im Grund 12, D-31867 Lauenau, Germany (e-mail:ludwig@hothorn.de)\\ \scriptsize(retired from Leibniz University Hannover)\\
\normalsize Mario Hasler,\\
  \scriptsize Applied Statistics, Faculty of Agricultural and Nutritional Sciences, Christian-Albrechts-University, Kiel, Germany}

\title{The Dunnett procedure with possibly heterogeneous variances}
\begin{document}

\maketitle
\begin{abstract}
Most comparisons of treatments or doses against a control are performed by the original Dunnett single step procedure \cite{Dunnett1955} providing both adjusted $p$-values and simultaneous confidence intervals for differences to the control. Motivated by power arguments, unbalanced designs with higher sample size in the control are recommended. When higher variance occur in the treatment of interest or in the control, the related per-pairs power is reduced, as expected. However, if the variance is increased in a non-affected treatment group, e.g. in the highest dose (which is highly significant), the per-pairs power is also reduced in the remaining treatment groups of interest. I.e., decisions about the significance of certain comparisons may be seriously distorted. To avoid this nasty property, three modifications for heterogeneous variances are compared by a simulation study with the original Dunnett procedure. For small and medium sample sizes, a Welch-type modification can be recommended. For medium to high sample sizes, the use of a sandwich estimator instead of the common mean square estimator is useful. Related CRAN packages are provided.\\
Summarizing we recommend not to use the original Dunnett procedure in routine and replace it by a robust modification. Particular care is needed in small sample size studies.
\end{abstract}


\section{Introduction}
\label{sec1}

Both clinical multi-arm trials, e.g. dose finding phase IIb studies, and non-clinical bioassays commonly use a placebo or zero-dose control for the comparisons against treatment or dose groups. Commonly, the  original Dunnett single step procedure \cite{Dunnett1955} is used. The question arises how robust is this procedure in the case of variance heterogeneity and still normally distributed errors. Several modifications are available, where primarily the summarizing concept of the any-pairs power (i.e., per-pairs power under $H_0$) was used to characterize the different power losses and primarily the control of the familywise error rate (FWER).\\

Summarizing, Dunnett's original test is conservative when low variances occur in groups with large sample size (with related power loss), but it is unacceptably liberal when high variances occur in treatments with small sample size (with seemingly, unacceptable power increase). Appropriate modifications control the FWER at the price of a power loss compared to the unacceptable power of the original under these conditions.


\section{A motivating example}
\label{sec2}

The serum creatine kinase content in the clinical chemistry raw data in rats after treatment with sodium dichromate are used as a motivating example \cite{Hothorn2016}. In this balanced design, both the low 62.5 dose and the high 1000 mg dose reveal higher variances.
\begin{figure}[htbp]
	\centering
		\includegraphics[width=0.4\textwidth]{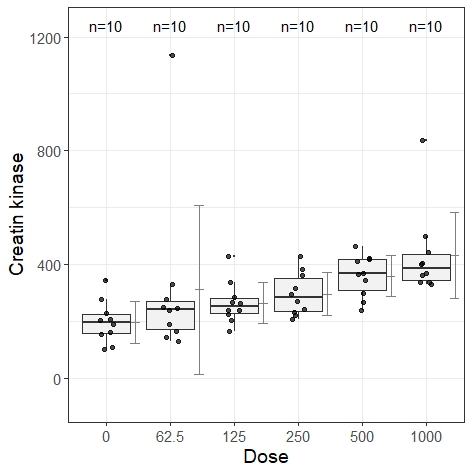}
	\caption{Creatin kinase example}
	\label{fig:Creatin}
\end{figure}

\begin{table}[ht]
\centering\small
	\begin{tabular}{l|l|l}
  \hline
 Comparison & $p$-value Dunnett original& $p$-value modified \\ 
  \hline
62.5 - 0 &  0.154 & 0.406\\ 
  125 - 0 & 0.407 & 0.108\\ 
  250 - 0 & 0.221 & 0.019\\ 
  500 - 0 & 0.036 & 0.0002\\ 
  1000 - 0 & 0.002 & 0.0017\\ 
   \hline
\end{tabular}
\caption{Adjusted $p$-values for comparisons against control (0)}
	\label{tab:Creatin}
\end{table}	
Whereas, the original Dunnett test reveals the 250 mg dose as not significant, the Welch-type modified approach reveals 'correctly' the 250 mg dose as significantly increased. The original test uses an increased global variance estimate by the high variances in the 62.5 and 1000 mg doses accordingly for all comparisons against control. Both are irrelevant for the inference of the 250 mg group.


\section{Alternative approaches robust against variance heterogeneity}
\label{sec5}

Three alternative approaches are considered here: i) the use of a sandwich estimator instead of the common variance estimator \cite{Herberich2010}, ii) the use of pairwise-specific Welch-type degree of freedom with related approximation of the multivariate $t$-distribution \cite{Hasler2008} and iii) Bonferroni-adjusted Welch-t tests. The last one is limited to a small number of treatments $k$ because it ignores the correlations between the marginal tests.

Dunnett's procedure can be formulated as a multiple contrast approach. A single contrast test is defined to 
 $	t_{Contrast}=\sum_{i=0}^k c_i\bar{y}_i/MQ_R \sqrt{\sum_i^k c_i^2/n_i}$ where $c_i$ are the contrast coefficients. The condition $\sum_{i=0}^k c_i=0$ guarantees a $t_{df,1-\alpha}$ distributed level-$\alpha$-test and compatible simultaneous confidence intervals requires further $\sum sign^+(c_{i})=1, \sum sign^-(c_{i})=1$.  The multiple contrast test is a maxT test: $t_{MCT}=max(t_1,...,t_q)$ whereas  $(t_1,\ldots,t_q)^\prime$ follows jointly a $q$-variate $t$-distribution with the common degree of freedom $df$ and a correlation matrix  $\textbf{R}$ ($\textbf{R}=f(c_{ij},n_i)$). The Dunnett contrasts coefficients are $-1$ in the control, $+1$ in the treatment of interest and $0$ otherwise. I.e. for a layout with $k+1$ groups, $k$ multiple contrasts are needed for one-sided comparisons and $2k$ for two-sided comparisons. Notice, this approach based on a common variance estimator $MQ_R$ and the common $df$ in this $k+1$ one-way layout. The approach ii) use pairwise specific Welch-type degree of freedom $df_{0,i}$ and a related approximation of the multivariate t-distribution \cite{Hasler2008}. A further modification \cite{Herberich2010} use of a sandwich estimator, a robust covariance matrix estimator which is heteroscedastic consistent (HC) \cite{Zeileis2006}, instead of the common $MQ_R$ estimator and still the common $df$.

\section{Simulation study}
Normally distributed variables with heterogeneous variances are considered in both a balanced and an unbalanced ($n_0>n_i$) $[k=3+1]$-design for small and moderate sample sizes. Under the null hypothesis $H_0$, the empirical size and the elementary $\alpha_i$ are estimated and under the alternative hypothesis $H_1$ the related any-pairs and per-pair power for 5000/2000 runs. Violation of \bl{size}, \gr{fair per-power reduction due to variance increase in exactly the considered group}, \re{distorted per-power reduction due to variance increase increase in a different group} are highlighted in the tables below.


\subsection{Small sample sizes}

For the original Dunnett test \cite{Dunnett1955} (Duo,$d_i$), sandwich estimator modification \cite{Herberich2010} (DuS,$S_i$), the PI method \cite{Hasler2008} (DuH, $h_i$) and the Bonferroni-Welch-tests (W0,$w_i$) both any-pairs power (FWER and $H_0$) and per-pairs power were estimated.


\subsubsection{Under $H_0$}

\begin{table}[ht]
\centering
\scalebox{0.6}{
\begin{tabular}{rr|rrrr||rrrr|rrrr|rrrr|rrrr}
  \hline
 $n_1$ & $n_i$ & $s_1$ & $s_2$ & $s_3$ & $s_4$ & Du0 & d1 & d2 & d3 & DuS & S1 & S2 & S3 & DuH & h1 & h2 & h3 & W0 & w1 & w2 & w3 \\ 
  \hline
 6 & 6 & 1 & 1 & 1 & 1 & 0.049 & 0.017 & 0.021 & 0.022 & 0.051 & 0.019 & 0.020 & 0.021 & 0.049 & 0.018 & 0.019 & 0.021 & 0.043 & 0.015 & 0.017 & 0.018 \\ 
 6 & 6 & 1 & 1 & 1 & 4 & \bl{0.080} & 0.000 & 0.000 & 0.080 & \bl{0.062} & 0.019 & 0.020 & 0.027 & 0.052 & 0.018 & 0.019 & 0.020 & 0.047 & 0.016 & 0.017 & 0.017 \\ 
   \hline
 9 & 5 & 1 & 1 & 1 & 1 & 0.057 & 0.021 & 0.022 & 0.023 & 0.054 & 0.019 & 0.020 & 0.021 & 0.054 & 0.018 & 0.020 & 0.022 & 0.048 & 0.016 & 0.018 & 0.019 \\ 
 9 & 5 & 1 & 1 & 1 & 4 &\bl{0.102} & 0.001 & 0.001 & 0.101 & \bl{0.061} & 0.018 & 0.018 & 0.028 & 0.052 & 0.016 & 0.018 & 0.019 & 0.048 & 0.015 & 0.016 & 0.018 \\ \hline
   \hline
\end{tabular}
}
\end{table}
In the homoscedastic case, all procedures control FWER, whereas the Bonferroni-Welch-test are conservative per construction. In the heterogeneous case, the original Dunnett test is unacceptably liberal and hence the power estimations is inadequate. In the partially unbalanced design, it is even more liberal. (Notice, the test is conservative when lower variances are in groups with larger $n_i$; not shown here). The sandwich estimator modification is asymptotic and therefore liberal for small $n_i$. All Welch-type approaches control $\alpha$ by means of reduced $df$ and hence they are conservative with an inherent related power loss (see below). 

\subsubsection{Under specific $H_1$}
Three blocks are considered with the minimum effective dose $MED=\mu_3, MED=\mu_2, MED=\mu_1$ with homogeneous and heterogeneous variances in a balanced and partial unbalanced design. The power estimations of the original Dunnett test are not comparable in the heterogeneous case because of its liberal behavior.\\
\textbf{Balanced design}\\
\begin{table}[ht]
\centering
\scalebox{0.55}{
\begin{tabular}{rrrr|rr||rrrr|rrrr|rrrr|rrrr|rrrr}
  \hline
$\mu_1$ &$\mu_2$ & $\mu_3$ & $\mu_4$ & $n_1$ & $n_i$ & $s_1$ & $s_2$ & $s_3$ & $s_4$ & Du0 & d1 & d2 & d3 & DuS & S1 & S2 & S3 & DuH & h1 & h2 & h3 & W0 & w1 & w2 & w3 \\ 
  \hline
5 & 5 & 5 & 3 & 6 & 6 & 1 & 1 & 1 & 1 & 0.890 & 0.020 & 0.017 & 0.890 & 0.854 & 0.019 & 0.015 & 0.853 & 0.846 & 0.018 & 0.015 & 0.845 & 0.817 & 0.014 & 0.012 & 0.816 \\ 
5 & 5 & 5 & 3 & 6 & 6 & 1 & 1 & 1 & 4 & 0.354 & 0.000 & 0.000 & \gr{0.354} & 0.202 & 0.018 & 0.017 & \gr{0.179} & 0.174 & 0.018 & 0.017 & \gr{0.150} & 0.164 & 0.018 & 0.014 & \gr{0.142} \\ 
5 & 5 & 5 & 3 & 6 & 6 & 1 & 1 & 4 & 1 & 0.264 & 0.000 & 0.070 & \re{0.228} & 0.839 & 0.023 & 0.028 & 0.836 & 0.836 & 0.022 & 0.020 & 0.833 & 0.814 & 0.020 & 0.018 & 0.812 \\ 
5 & 5 & 5 & 3 & 6 & 6 & 1 & 4 & 1 & 1 & 0.270 & 0.070 & 0.001 & \re{0.237} & 0.826 & 0.027 & 0.015 & 0.824 & 0.821 & 0.022 & 0.013 & 0.818 & 0.806 & 0.021 & 0.011 & 0.802 \\ 
5 & 5 & 5 & 3 & 6 & 6 & 4 & 1 & 1 & 1 & 0.366 & 0.068 & 0.068 & \gr{0.366} & 0.234 & 0.035 & 0.035 & \gr{0.234} & 0.207 & 0.031 & 0.029 & \gr{0.207} & 0.130 & 0.016 & 0.019 & \gr{0.130} \\ \hline
5 & 5 & 3 & 3 & 6 & 6 & 1 & 1 & 1 & 1 & 0.958 & 0.017 & 0.892 & 0.879 & 0.930 & 0.015 & 0.840 & 0.835 & 0.929 & 0.016 & 0.837 & 0.827 & 0.905 & 0.012 & 0.807 & 0.794 \\ 
5 & 5 & 3 & 3 & 6 & 6 & 1 & 1 & 1 & 4 & 0.480 & 0.000 & \re{0.238} & 0.380 & 0.862 & 0.022 & 0.841 & \gr{0.180} & 0.851 & 0.022 & 0.833 & 
\gr{0.141} & 0.835 & 0.019 & 0.818 & \gr{0.130} \\ 
5 & 5 & 3 & 3 & 6 & 6 & 1 & 1 & 4 & 1 & 0.464 & 0.000 & \gr{0.363} & \re{0.224} & 0.857 & 0.020 & \gr{0.168} & 0.838 & 0.846 & 0.019 & \gr{0.128} & 0.831 & 0.829 & 0.018 & \gr{0.121} & 0.811 \\ 
5 & 5 & 3 & 3 & 6 & 6 & 1 & 4 & 1 & 1 & 0.353 & 0.076 & \re{0.239} & \re{0.235} & 0.932 & 0.030 & 0.833 & 0.827 & 0.927 & 0.021 & 0.822 & 0.823 & 0.916 & 0.018 & 0.804 & 0.808 \\ 
5 & 5 & 3 & 3 & 6 & 6 & 4 & 1 & 1 & 1 & 0.403 & 0.068 & \gr{0.353} & 0.363 & 0.266 & 0.031 & \gr{0.226} & 0.230 & 0.230 & 0.025 & \gr{0.192} & 0.198 & 0.142 & 0.016 & \gr{0.121} & 0.121 \\ 
\hline
 5 & 3 & 3 & 3 & 6 & 6 & 1 & 1 & 1 & 1 & 0.984 & 0.896 & 0.908 & 0.893 & 0.974 & 0.865 & 0.858 & 0.852 & 0.974 & 0.857 & 0.854 & 0.847 & 0.960 & 0.826 & 0.828 & 0.818 \\ 
5 & 3 & 3 & 3 & 6 & 6 & 1 & 1 & 1 & 4 & 0.514 & \re{0.258} & 0.238 & 0.359 & 0.938 & 0.830 & 0.846 & \gr{0.182} & 0.931 & 0.820 & 0.839 & 
\gr{0.145} & 0.919 & 0.801 & 0.818 & \gr{0.135} \\ 
 5 & 3 & 3 & 3 & 6 & 6 & 1 & 1 & 4 & 1 & 0.502 & \re{0.225} & 0.355 & 0.225 & 0.932 & 0.824 & \gr{0.173} & 0.816 & 0.927 & 0.816 & \gr{0.135} & 0.811 & 0.915 & 0.797 & \gr{0.128} & 0.790 \\ 
 5 & 3 & 3 & 3 & 6 & 6 & 1 & 4 & 1 & 1 & 0.532 & \gr{0.378} & 0.249 & 0.235 & 0.936 & \gr{0.189} & 0.829 & 0.829 & 0.930 & \gr{0.146} & 0.821 & 0.820 & 0.917 & \gr{0.136} & 0.804 & 0.803 \\ 
 5 & 3 & 3 & 3 & 6 & 6 & 4 & 1 & 1 & 1 & 0.410 & \gr{0.355} & 0.350 & 0.346 & 0.282 & \gr{0.230} & 0.227 & 0.230 & 0.243 & \gr{0.204} & 0.193 & 0.196 & 0.155 & \gr{0.126} & 0.124 & 0.128 \\ 
\hline\hline
\end{tabular}
}
\end{table}
In all tests we observe a fair, remarkable power loss when the increase variance occur in the control (hence for all comparisons) and the intended MED group (in \gr{green}). The less remarkable power loss in the sandwich approach is due to its liberal behavior for small sample sizes
The power loss of the PI approach is less strong compared to Bonferroni per- definition. The most important findings are the power loss in the non-intended groups in the original Dunnett test (in \re{red}). E.g. $MED=\mu_3, s_4=s_1=s_2=s_i; s_3>s_i$. Compared with the modified approaches this per-pairs power is remarkable, i.e. it reveals a distort MED estimation. This power loss is more in the balanced design.\\

\textbf{Unbalanced design}\\

\begin{table}[ht]
\centering
\scalebox{0.55}{
\begin{tabular}{rrrr|rr||rrrr|rrrr|rrrr|rrrr|rrrr}
  \hline
$\mu_1$ &$\mu_2$ & $\mu_3$ & $\mu_4$ & $n_1$ & $n_i$ & $s_1$ & $s_2$ & $s_3$ & $s_4$ & Du0 & d1 & d2 & d3 & DuS & S1 & S2 & S3 & DuH & h1 & h2 & h3 & W0 & w1 & w2 & w3 \\ 
  \hline

  5 & 5 & 5 & 3 & 9 & 5 & 1 & 1 & 1 & 1 & 0.909 & 0.019 & 0.018 & 0.909 & 0.851 & 0.021 & 0.019 & 0.850 & 0.828 & 0.020 & 0.021 & 0.827 & 0.812 & 0.019 & 0.015 & 0.812 \\ 
  5 & 5 & 5 & 3 & 9 & 5 & 1 & 1 & 1 & 4 & 0.430 & 0.002 & 0.000 & \gr{0.430} & 0.194 & 0.020 & 0.012 & \gr{0.171} & 0.147 & 0.019 & 0.012 & \gr{0.121} & 0.139 & 0.018 & 0.011 & \gr{0.114} \\ 
  5 & 5 & 5 & 3 & 9 & 5 & 1 & 1 & 4 & 1 & 0.405 & 0.002 & 0.113 & \re{0.350} & 0.844 & 0.024 & 0.026 & 0.839 & 0.814 & 0.022 & 0.019 & 0.809 & 0.803 & 0.021 & 0.018 & 0.799 \\ 
  5 & 5 & 5 & 3 & 9 & 5 & 1 & 4 & 1 & 1 & 0.394 & 0.110 & 0.002 & \re{0.340} & 0.839 & 0.034 & 0.018 & 0.835 & 0.811 & 0.023 & 0.015 & 0.807 & 0.801 & 0.021 & 0.015 & 0.797 \\ 
  5 & 5 & 5 & 3 & 9 & 5 & 4 & 1 & 1 & 1 & 0.234 & 0.022 & 0.018 & \gr{0.234} & 0.280 & 0.028 & 0.028 & \gr{0.280} & 0.278 & 0.028 & 0.030 & \gr{0.278} & 0.200 & 0.019 & 0.014 & \gr{0.200} \\ 
\hline
 5 & 5 & 3 & 3 & 9 & 5 & 1 & 1 & 1 & 1 & 0.974 & 0.019 & 0.897 & 0.903 & 0.960 & 0.018 & 0.855 & 0.850 & 0.952 & 0.020 & 0.830 & 0.828 & 0.942 & 0.016 & 0.814 & 0.809 \\ 
 5 & 5 & 3 & 3 & 9 & 5 & 1 & 1 & 1 & 4 & 0.568 & 0.002 & \re{0.333} & 0.421 & 0.862 & 0.020 & 0.845 & 0.163 & 0.834 & 0.021 & 0.819 & 0.108 & 0.823 & 0.021 & 0.808 & 0.102 \\ 
 5 & 5 & 3 & 3 & 9 & 5 & 1 & 1 & 4 & 1 & 0.568 & 0.002 & \gr{0.412} & 0.342 & 0.859 & 0.019 & \gr{0.160} & 0.837 & 0.826 & 0.018 & \gr{0.116} & 0.807 & 0.818 & 0.018 & \gr{0.111} & 0.800 \\ 
 5 & 5 & 3 & 3 & 9 & 5 & 1 & 4 & 1 & 1 & 0.482 & 0.117 & \re{0.336} & 0.329 & 0.952 & 0.037 & 0.829 & 0.833 & 0.944 & 0.024 & 0.797 & 0.818 & 0.940 & 0.021 & 0.789 & 0.808 \\
 5 & 5 & 3 & 3 & 9 & 5 & 4 & 1 & 1 & 1 & 0.289 & 0.024 & \gr{0.244} & 0.243 & 0.349 & 0.030 & \gr{0.297} & 0.295 & 0.351 & 0.031 & \gr{0.300} & 0.296 & 0.252 & 0.020 & \gr{0.210} & 0.208 \\  \hline

 5 & 3 & 3 & 3 & 9 & 5 & 1 & 1 & 1 & 4 & 0.623 & \re{0.338} & 0.330 & 0.415 & 0.959 & 0.831 & 0.844 & 0.160 & 0.950 & 0.807 & 0.818 & 0.105 & 0.947 & 0.803 & 0.808 & 0.103 \\ 
 5 & 3 & 3 & 3 & 9 & 5 & 1 & 1 & 4 & 1 & 0.639 & \re{0.358} & 0.420 & 0.343 & 0.966 & 0.859 & 0.156 & 0.848 & 0.956 & 0.824 & 0.102 & 0.818 & 0.952 & 0.814 & 0.096 & 0.814 \\ 
 5 & 3 & 3 & 3 & 9 & 5 & 1 & 4 & 1 & 1 & 0.636 & \gr{0.422} & 0.331 & 0.336 & 0.964 & \gr{0.163} & 0.849 & 0.844 & 0.950 & \gr{0.116} & 0.809 & 0.817 & 0.950 & \gr{0.111} & 0.802 & 0.810 \\ 
 5 & 3 & 3 & 3 & 9 & 5 & 4 & 1 & 1 & 1 & 0.310 & \gr{0.234} & 0.237 & 0.235 & 0.365 & \gr{0.286} & 0.291 & 0.282 & 0.366 & \gr{0.289} & 0.295 & 0.281 & 0.272 & \gr{0.201} & 0.205 & 0.203 \\ 
   \hline
\end{tabular}
}
\end{table}


\subsection{Moderate sample sizes}
\subsubsection{Under $H_0$}

\begin{table}[ht]
\centering
\scalebox{0.55}{
\begin{tabular}{rrrrrr|rrrr||rrrr|rrrr|rrrr|rrrr}
  \hline
 $\mu_1$ &$\mu_2$ & $\mu_3$ & $\mu_4$ & $n_1$ & $n_i$ & $s_1$ & $s_2$ & $s_3$ & $s_4$ & Du0 & d1 & d2 & d3 & DuS & S1 & S2 & S3 & DuH & h1 & h2 & h3 & W0 & w1 & w2 & w3 \\ 
  \hline
5 & 5 & 5 & 5 & 20 & 20 & 1 & 1 & 1 & 1 & 0.051 & 0.020 & 0.020 & 0.020 & 0.049 & 0.018 & 0.020 & 0.020 & 0.052 & 0.019 & 0.021 & 0.020 & 0.045 & 0.016 & 0.019 & 0.018 \\ 
5 & 5 & 5 & 5 & 20 & 20 & 1 & 1 & 1 & 4 & \bl{0.064} & 0.000 & 0.000 & 0.064 & 0.050 & 0.017 & 0.019 & 0.018 & 0.050 & 0.017 & 0.020 & 0.017 & 0.046 & 0.016 & 0.019 & 0.015 \\ 
  \hline
\end{tabular}
}
\end{table}
The violation of FWER for variance heterogeneity is less pronounced for small sample sizes. The sandwich test controls FWER hereby.

\subsubsection{Under $H_1$}

\begin{table}[ht]
\centering
\scalebox{0.55}{
\begin{tabular}{rrrrrr|rrrr||rrrr|rrrr|rrrr|rrrr}
  \hline
 $\mu_1$ &$\mu_2$ & $\mu_3$ & $\mu_4$ & $n_1$ & $n_i$ & $s_1$ & $s_2$ & $s_3$ & $s_4$ & Du0 & d1 & d2 & d3 & DuS & S1 & S2 & S3 & DuH & h1 & h2 & h3 & W0 & w1 & w2 & w3 \\ 
	  \hline
 5 & 5 & 4 & 4 & 20 & 20 & 1 & 1 & 1 & 1 & 0.945 & 0.012 & 0.845 & 0.850 & 0.935 & 0.011 & 0.827 & 0.833 & 0.938 & 0.012 & 0.830 & 0.837 & 0.920 & 0.009 & 0.807 & 0.814 \\ 
 5 & 5 & 4 & 4 & 20 & 20 & 1 & 1 & 1 & 4 & 0.381 & 0.000 & \re{0.123} & 0.319 & 0.852 & 0.017 & 0.841 & 0.141 & 0.854 & 0.018 & 0.843 & 0.136 & 0.843 & 0.016 & 0.832 & 0.126 \\ 
 5 & 5 & 4 & 4 & 20 & 20 & 1 & 1 & 4 & 1 & 0.383 & 0.000 & \gr{0.327} & 0.112 & 0.852 & 0.015 & \gr{0.156} & 0.834 & 0.855 & 0.015 & \gr{0.146} & 0.838 & 0.843 & 0.015 & \gr{0.138} & 0.825 \\ \hline
 5 & 5 & 4 & 4 & 20 & 20 & 1 & 4 & 1 & 1 & 0.229 & 0.068 & \re{0.122} & 0.117 & 0.936 & 0.023 & 0.841 & 0.831 & 0.937 & 0.020 & 0.844 & 0.834 & 0.930 & 0.018 & 0.832 & 0.821 \\ 
 5 & 5 & 4 & 4 & 20 & 20 & 4 & 1 & 1 & 1 & 0.380 & 0.066 & \gr{0.328} & 0.335 & 0.269 & 0.035 & \gr{0.227} & 0.227 & 0.262 & 0.035 & \gr{0.222} & 0.218 & 0.165 & 0.018 & \gr{0.136} & 0.130 \\ 

   \hline
\end{tabular}
}
\end{table}
Even for moderate sample sizes, the power loss in the unaffected treatment groups is remarkable in the original Dunnett test whereas the power in the modified tests is unaffected.


\section{Summary}
\label{sec14}
\normalsize
The Dunnett procedure is recently also used in small sample size studies such as $n_i=3$ \cite{Hassaneen2022} or $n_i=5$ \cite{Satomoto2022}. Because  variance heterogeneity is likely in these bio-medical studies, the original Dunnett procedure should not be used and replaced by robust modifications, such as a Welch-type \cite{Hasler2008}. Under these data conditions, the usual Dunnett procedure may violate the FWER and reveal an unacceptable power loss in the unaffected treatment groups. From a practical point of view, one should not be so much concerned about liberal behavior and a power loss in this context, but about the possibility of distort treatment identification. Simply, one cannot identify a low effective dose as significant. But this is to be avoided. \\

Particularly for small sample sizes under variance homogeneity, the robust modifications reveal a related power loss, both in any-pairs power and per-pairs power. For moderate and large sample sizes, the sandwich-type modification represents an alternative approach \cite{Herberich2010}.
Still we recommend the modified procedures instead of the original Dunnett procedure in practice. We recommend to tolerate a small power loss in the variance homogeneous case to avoid the more critical distortion in the variance heterogeneous case.\\
Further research is directed to ratio-to-control as effect size and endpoints in the generalized linear model, such as counts.


\section{Appendix R-Code}

\scriptsize
\begin{verbatim}
myC <-
  structure(list(Dose = c(0, 0, 0, 0, 0, 0, 0, 0, 0, 0, 62.5, 62.5, 
                          62.5, 62.5, 62.5, 62.5, 62.5, 62.5, 62.5, 62.5, 125, 125, 125, 
                          125, 125, 125, 125, 125, 125, 125, 250, 250, 250, 250, 250, 250, 
                          250, 250, 250, 250, 500, 500, 500, 500, 500, 500, 500, 500, 500, 
                          500, 1000, 1000, 1000, 1000, 1000, 1000, 1000, 1000, 1000, 1000
  ), CreatKinase = c(202, 205, 188, 155, 160, 229, 107, 101, 277, 
                     343, 240, 276, 247, 164, 144, 1135, 131, 189, 250, 330, 336, 
                     428, 239, 265, 239, 284, 165, 203, 224, 264, 384, 220, 206, 271, 
                     241, 362, 295, 317, 429, 233, 369, 366, 462, 409, 237, 419, 268, 
                     344, 297, 420, 403, 336, 330, 362, 497, 444, 401, 337, 838, 370
  ), dose = structure(c(1L, 1L, 1L, 1L, 1L, 1L, 1L, 1L, 1L, 1L, 
                        2L, 2L, 2L, 2L, 2L, 2L, 2L, 2L, 2L, 2L, 3L, 3L, 3L, 3L, 3L, 3L, 
                        3L, 3L, 3L, 3L, 4L, 4L, 4L, 4L, 4L, 4L, 4L, 4L, 4L, 4L, 5L, 5L, 
                        5L, 5L, 5L, 5L, 5L, 5L, 5L, 5L, 6L, 6L, 6L, 6L, 6L, 6L, 6L, 6L, 
                        6L, 6L), levels = c("0", "62.5", "125", "250", "500", "1000"), 
												class = "factor")), row.names = c(NA,60L), class = "data.frame")

library(toxbox); library(SimComp); library(multcomp)
boxclust(data=myC, outcome="CreatKinase", treatment="Dose",  ylabel="Creatin kinase",  xlabel="Dose", 
        option="uni", hjitter=0.125, legpos ="none", printN="FALSE", white=TRUE, psize=1.5, vlines="bg")
mod2<-lm(CreatKinase~dose, data=myC)
summary(glht(mod2, linfct = mcp(dose = "Dunnett"), alternative="greater")) # original Dunnett test
SimTestDiff(data=myC, grp="dose", resp="CreatKinase",
            type="Dunnett", alternative="greater", covar.equal=FALSE) # Welch-df- modified test
\end{verbatim}


\footnotesize
\bibliographystyle{plain}


\end{document}